\documentclass[doublespacing]{elsart}

\usepackage{graphicx}
\usepackage{amssymb}

\begin{document}
PC-2/ISS~2001\hfill\par
Submitted 17 October, 2001\hfill\par

\begin{frontmatter}


\journal{PC-2/ISS~2001: Version 1}

\title{Energy bandstructures of MgB$_2$ and the related compounds}

%
%
%
%
%
%

\author{Hisatomo Harima}

%
 
\address{The Institute of Scientific and Industrial Research, Osaka University, Ibaraki, Osaka 567-0047, Japan}

\begin{abstract}

Energy band structures are calculated for the new superconductor MgB$_2$ and the related compounds by using the LDA and an FLAPW method. 
It is found that the strong three dimensional network in low-lying $\pi$ bands brings about two dimensional $\sigma$ holes in MgB$_2$, which should be responsible for the superconductivity.
The de Haas-van Alphen frequencies and the cyclotron masses are obtained for the Fermi surfaces.
The $\sigma$ hole is not found in LiBC due to the less three dimensional network.
MgB$_2$C$_2$ is found a semiconductor, but the top of the valence bands are similar to the $\sigma$ bands in MgB$_2$.
The total energy calculations are also performed for Mg$_{1-x}$Al$_x$B$_2$ to investigate the structural phase instability, experimentally observed around $x=$10\%.

\

PACS codes:71.18.+y, 71.20.-b, 74.25, 74-70.-b
\end{abstract}

%
%

\begin{keyword}

band structure \sep MgB$_2$ \sep Fermi surface

\end{keyword}


\end{frontmatter}

%
%
%
%
%
\newpage
\section{Introduction} 

From the discovery of the superconductivity in MgB$_2$\cite{nagamatsu},
a lot of theoretical and experimental researches have been performed
in order to reveal the mechanism of the really high superconducting transition temperature ($T_{\rm sc}$) in metals and find a route to a new superconductor with much higher $T_{\rm sc}$.
It has been indicated that MgB$_2$ is a phonon-mediated BCS s-wave superconductor by theoretical calculations\cite{band1,band2},  the NMR study\cite{kitaoka} and the high-resolution photoemission study\cite{takahashi},
then detailed phonon dispersion is calculated\cite{phonon1,phonon2}.
On the other hand, multiple superconducting gaps are also indicated by the specific heat measurement\cite{hinetsu}, the tunneling spectroscopy study\cite{tunnel1,tunnel2} and the high-resolution photoemission study\cite{shin}.
In the electron doped system Mg$_{1-x}$Al$_x$B$_2$, the structural instabilityhas been observed around $x$ 10\%, resulting in the loss of superconductivity.\cite{mgalb2}

In this paper, the Fermi surface property in MgB$_2$ is theoretically obtained for the further microscopic measurements of single crystals.
The bandstructures for iso-electronic compounds LiBC and MgB$_2$C$_2$ are calculated for the comparison.
Total energy calculations for Mg$_{1-x}$Al$_x$B$_2$ are performed to clarify the structural instability.

\section{Method of Calculation} 

Band structure calculations are carried out by using 
a full potential APW (FLAPW) method
with the local density approximation (LDA) for the exchange correlation 
potential.
For the LDA, the formula proposed by Gunnarsson and
Lundqvist\cite{Gunnarsson1976} is used.
For the band structure calculation, we used the program codes;
TSPACE\cite{yanase1995} and KANSAI-99.

We used in the calculations the experimentally observed lattice constants: {\bf a}= 3.083\AA\ and {\bf c}=3.527\AA\ for MgB$_{2}$.
Muffin-tin (MT) radii are set as 0.4979 {\bf a} and 0.2742 {\bf a} for Mg and B, respectively.
Core electrons (Ne-core for Mg, He-core for B) 
are calculated inside the MT spheres in each 
self-consistent step.
The LAPW basis functions are truncated at $|{\bf k} + {\bf G}_i|
\le 3.33 \times 2\pi/{\bf a}$,
corresponding to 225 LAPW functions at the $\Gamma$ point.
The sampling points are uniformly distributed in the irreducible 1/24 th of 
the Brillouin zone,
333 {\bf k}-points (divided by 12, 12, 12) for both the potential convergence and the final band structure.
The similar parameters are used in calculations for LiBC and Mg$_{1-x}$Al$_x$B$_2$.

MgB$_2$C$_2$ crystallizes in a base centered orthorhombic structure containing 8 formula units in the primitive unit cell.
In this case, calculations are performed by using about 1,600 LAPW basis functions at the uniformly distributed 39 {\bf k}-points for the potential convergence and 369 {\bf k}-points (divided by 8, 8, 8) for the final band structure in the irreducible 1/8 th of the Brillouin zone.

\section{Bandstructure and Fermi surface of MgB$_2$} 

\begin{figure}[mgb2band]

     \centering
     \includegraphics{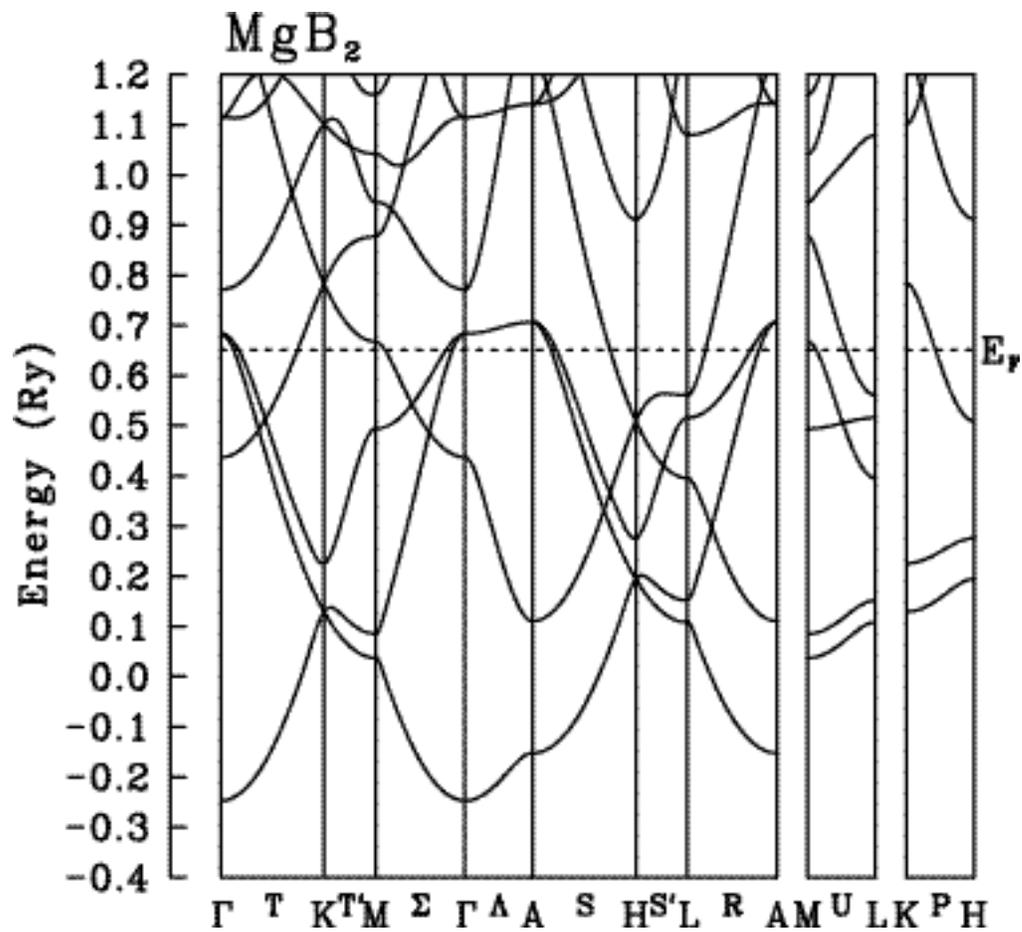}
     \caption{The calculated bandstructure for MgB$_{2}$.
} 
 \label{mgb2band}
 \end{figure}

\begin{figure}[mgb2dos]
     \centering
     \includegraphics{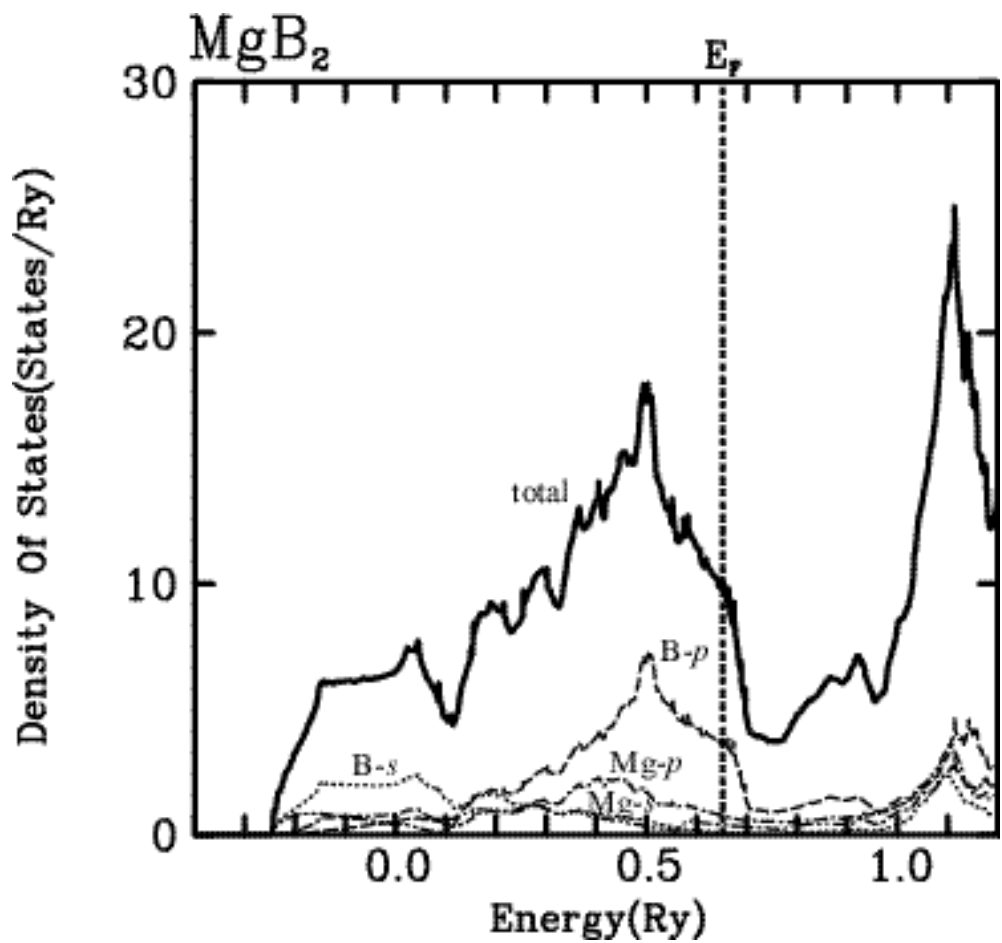}
     \caption{The calculated total and partial density of states for MgB$_{2}$.
} 
 \label{mgb2dos}
 \end{figure}

\begin{figure}[mgb2fermi]
     \centering
     \includegraphics{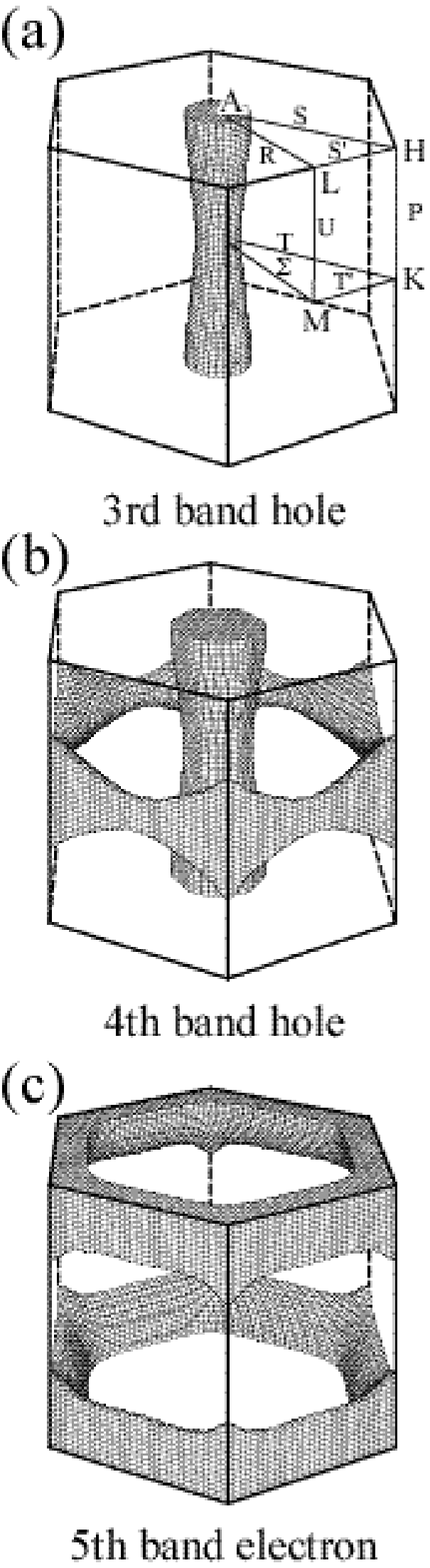}
     \caption{The calculated Fermi surfaces for MgB$_{2}$; (a) hole Fermi surface from the 3rd band, (b) hole Fermi surface from the 4th band and (c) electron Fermi surface from the 5th band.
} 
 \label{mgb2fermi}
 \end{figure}

The calculated energy bandstructure and the density of states are shown in Fig.~\ref{mgb2band} and Fig.~\ref{mgb2dos}, respectively.
They are very similar to the previous calculations\cite{band1,band2}, though Fig.~\ref{mgb2band} shows bandstructures along all the symmetry axes;
the $\Gamma$-K-M-$\Gamma$ lines is parallel to the A-H-L-A lines on the basal plane and the $\Lambda$, U and P axes are along the c-axis (see Fig.~\ref{mgb2fermi} (a)).

Let us see the dispersion along the P axis; the dispersive band crossing the Fermi level.
The band is originated from B-$p_z$($\pi$) band and doubly degenerated.
These B-$p_z$($\pi$) states are non-bonding along the P axis within the honeycomb lattice, and anti-bonding (bonding) at the K (H) point for the inter-layer (along the c-axis) mixing.
While B-$p_{xy}$($\sigma$) states are split to bonding and anti-bonding sates within the honeycomb lattice, 
then the top of the bonding states are located just above the Fermi level along the $\Lambda$ axis.
While the bottom of the anti-bonding bands are located around 1.1 Ry.
The energy and its components foe the states near the Fermi level are listed in Table~\ref{tablestates}.
Eventually, there are four Fermi surfaces;
two cylindrical hole Fermi surfaces along the $\Lambda$ axis originating in the $\sigma$ band,
and from the $\pi$ bands one hole around the T' axis one electron along S' axis, 
as shown in Fig.~\ref{mgb2fermi}.
The total number of holes and electrons are compensated.

The unique characteristics in the bandstructure of MgB$_2$ is the existence of the cylindrical $\sigma$-holes (Fig.~\ref{mgb2fermi} (a) and (b)).
The situation has been discussed as the relative shift of the $\sigma$ and $\pi$ bands\cite{band1}.
In fact, the totally non-bonding $\sigma$ states (about 0.91 Ry) is much higher in energy than the non-bonding $\pi$ state (0.65 Ry; the mid-point on the P axis).
It should be ascribed to the ionic Mg$^{2+}$ strongly affecting the crystalline field for the B-$p$ states.
Moreover, here it is emphasized that the 3D network in the $\pi$ band is suggested by the dispersive band along the P axis.

As was already discussed\cite{band1}, graphite (primitive graphite C$_2$) shows a 2D bandstructure.
In graphite, the inter-layer distance ($d_{layer}=$3.348\AA) is smaller than in MgB$_2$ ($d_{layer}=$3.528\AA),
but the planar bond length ($d_{bond}=$1.418 \AA) is much smaller than in MgB$_2$ ($d_{bond}=$1.780 \AA).
The planar bond length determines the total band width in each case, because of the much smaller distance and more coordination in plane,
though it depends on how extended the wave functions are.
As a result, the valence band width is narrower in MgB$_2$ than graphite.
However, what should be noticed here is the ratio of $d_{layer}$ and $d_{bond}$, which is 2.36 in graphite and 1.98 in MgB$_2$.
The much smaller ratio in MgB$_2$ causes more 3D feature, resulting in the dispersive $\pi$ band along the P axis, which obtains both the hole and electron Fermi surfaces.
The $\sigma$ bands keep the 2D character, because the $p_{xy}$ orbitals are extended in a layer.
Eventually, the strong 3D feature in low-lying $\pi$ bands brings about 2D Fermi surface of the $\sigma$ band in MgB$_2$, as shown in Fig.~\ref{mgb2fermi}.
The 3D feature and the low-lying in $\pi$ bands are ascribed to the existence of Mg$^{2+}$ ion.

The calculated specific heat coefficient $\gamma_{band}$ is 1.73 mJ/mol K$^2$, while the experimental value $\gamma_{exp}$ is 2.6 mJ/mol K$^2$.\cite{hinetsu}
Therefore, the averaged electron-phonon mass enhancement factor $\lambda$ is derived as 0.50 from $\gamma_{exp}=(1+\lambda) \gamma_{band}$.

\begin{figure}[mgb2dhva]
     \centering
     \includegraphics{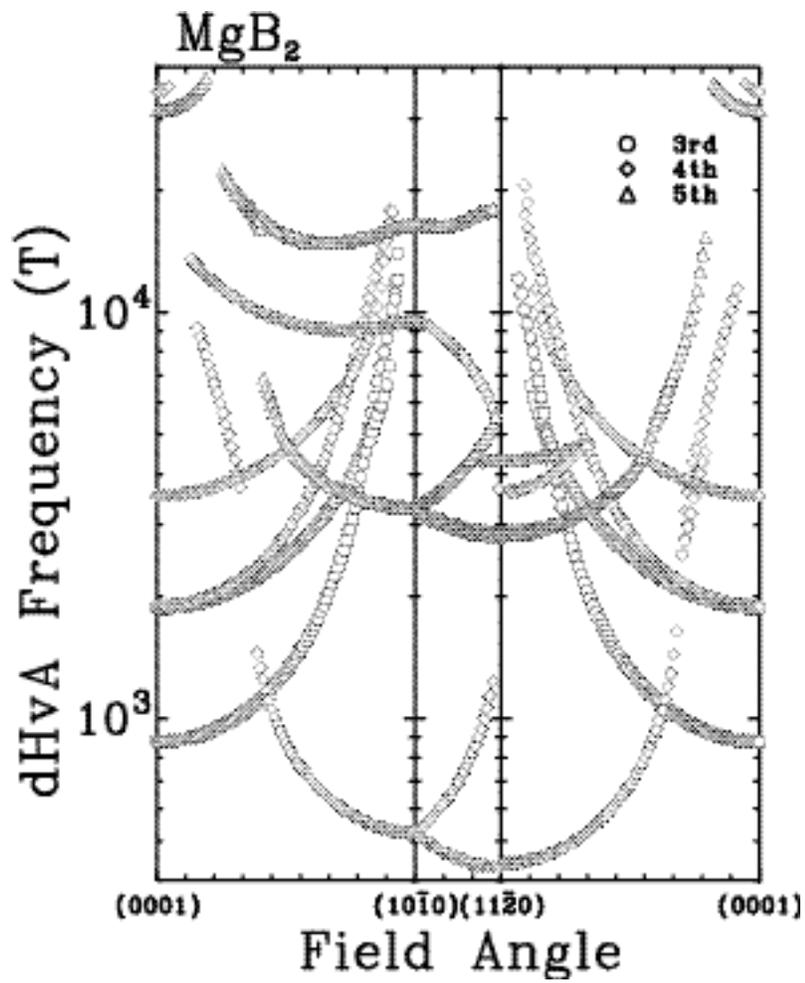}
     \caption{The calculated angular dependence od the dHvA frequencies for MgB$_{2}$.
} 
 \label{mgb2dhva}
 \end{figure}

The calculated angular dependence of the de Haas-van Alphen (dHvA) frequencies are shown in Fig.~\ref{mgb2dhva}.
The dHvA frequencies and the cyclotron masses are summarized in Table~\ref{tabledhva}.
Several theoretical studies have pointed out the importance of electron-phonon coupling in the $\sigma$ band for the superconductivity,
then the superconducting gap in MgB$_2$ can be expected to be larger on the $\sigma$ Fermi surfaces than on the $\pi$ Fermi surfaces.
It is also expected to be larger mass enhancement in the cyclotron masses in the $\sigma$ Fermi surfaces, though they have not yet observed so far, unfortunately.
Recently MgB$_2$ single crystals are grown by a several groups,\cite{singlelee,kitazawa,pohan}
then the dHvA oscillations is expected to be detected soon.
The experimentally observed cyclotron masses will be the direct evidence for the two superconducting gaps in MgB$_2$.

\section{Bandstructures of LiBC and MgB$_2$C$_2$} 

\begin{figure}[mgb2c2crys]
     \centering
     \includegraphics{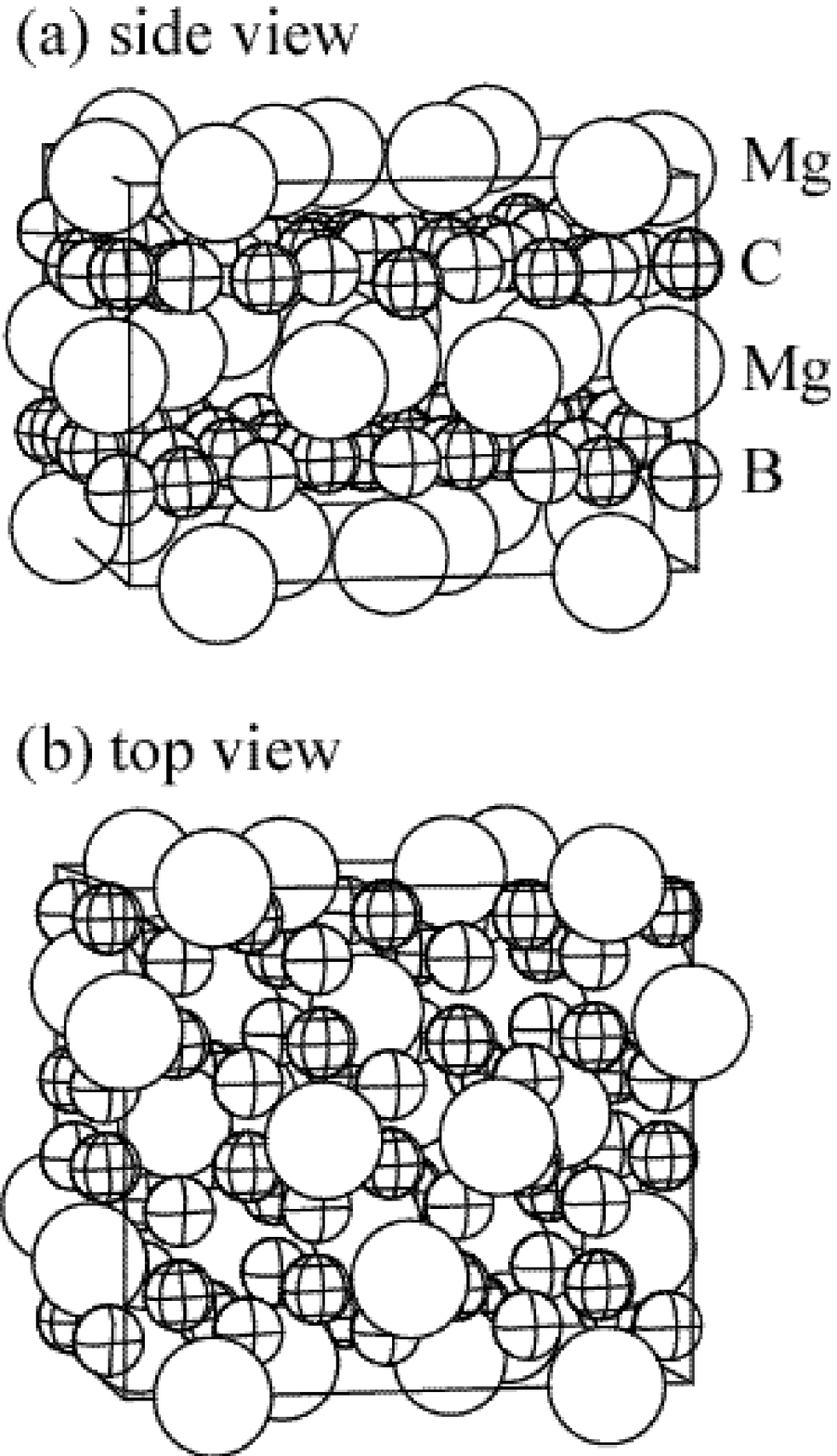}
     \caption{The crystal structure of MgB$_{2}$C$_{2}$; (a) side view and (b) top view.
} 
 \label{mgb2c2crys}
 \end{figure}

It is worth to see bandstructures of the related compounds.
LiBC and MgB$_2$C$_2$ are iso-electric compounds with MgB$_2$, though the crystal structure is slightly different.\cite{mgb2c2cry}
LiBC can be obtained, if Mg and one of B in MgB$_2$ are replaced by Li and C.
The stacking along the c-direction has not been experimentally determined, 
so we assume B-B-B...(C-C-C...) stacking.
MgB$_2$C$_2$ crystallizes in a orthorhombic structure, as shown in Fig.~\ref{mgb2c2crys}, in which there are a distorted B-C honeycomb lattice layer and a Mg layer.
In LiBC, $d_{layer}=$3.53 \AA, $d_{bond}=$1.59 \AA\ and $d_{layer}/d_{bond}=$2.22.
While, in MgB$_2$C$_2$, $d_{layer}=$3.74 \AA, $d_{bond}=$1.58 \AA\ in average and $d_{layer}/d_{bond}=$2.37.
Therefore, both compounds are expected to be less 3D than MgB$_2$.

\begin{figure}[libcband]
     \centering
     \includegraphics{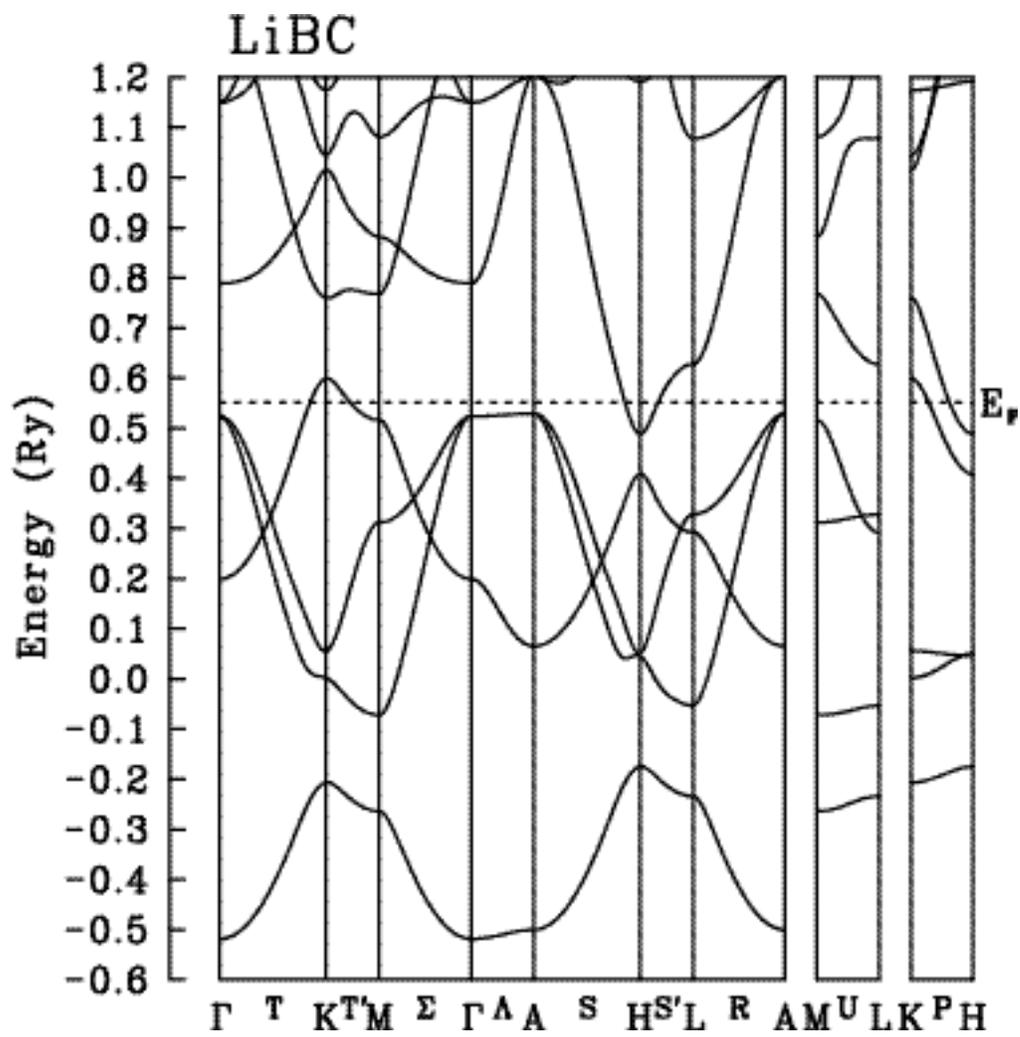}
     \caption{The calculated bandstructure for LiBC.
} 
 \label{libcband}
 \end{figure}

\begin{figure}[libcfermi]
     \centering
     \includegraphics{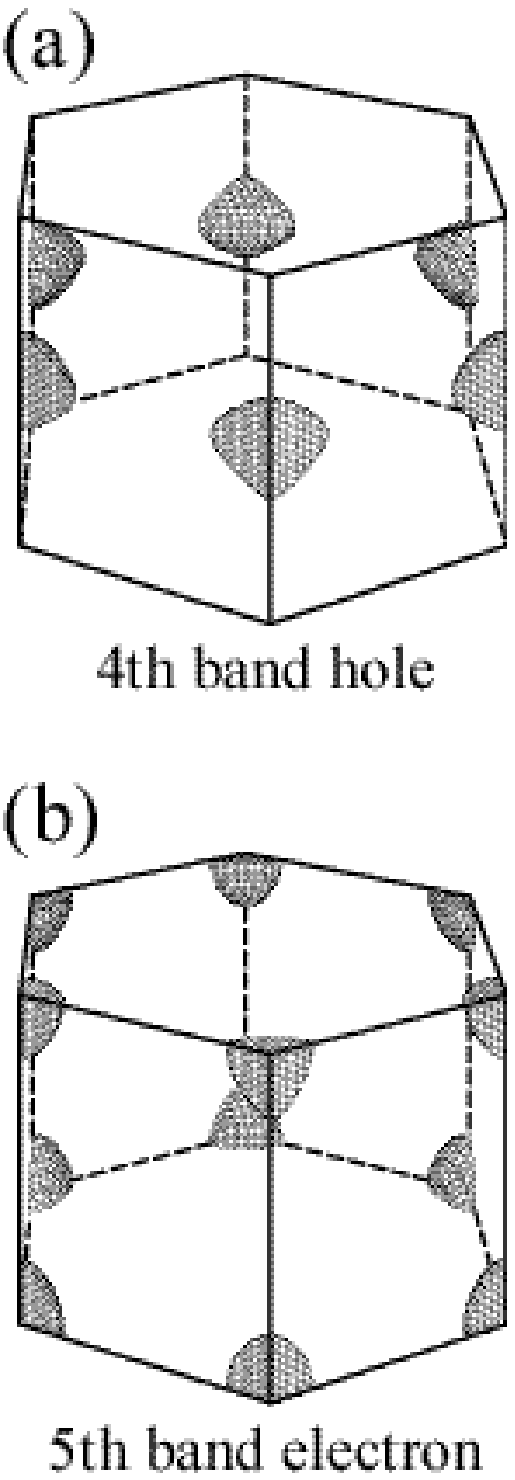}
     \caption{The calculated Fermi surfaces for LiBC; (a) hole Fermi surface from the 4th band and (b) electron Fermi surface from the 5th band.
} 
 \label{libcfermi}
 \end{figure}

Figure~\ref{libcband} shows the bandstructure for LiBC.
The space group of LiBC is P\=6$m$2, then the degeneracy along the P-axis is lifted.
Li$^{1+}$ ion has less attractive for the $\pi$ bands, then the bonding $\sigma$ bands get under the Fermi level.
Eventually, there are hole and electron Fermi surfaces both from the $\pi$ bands, as shown in Fig.~\ref{libcfermi}.

\begin{figure}[mgb2c2band]
     \centering
     \includegraphics{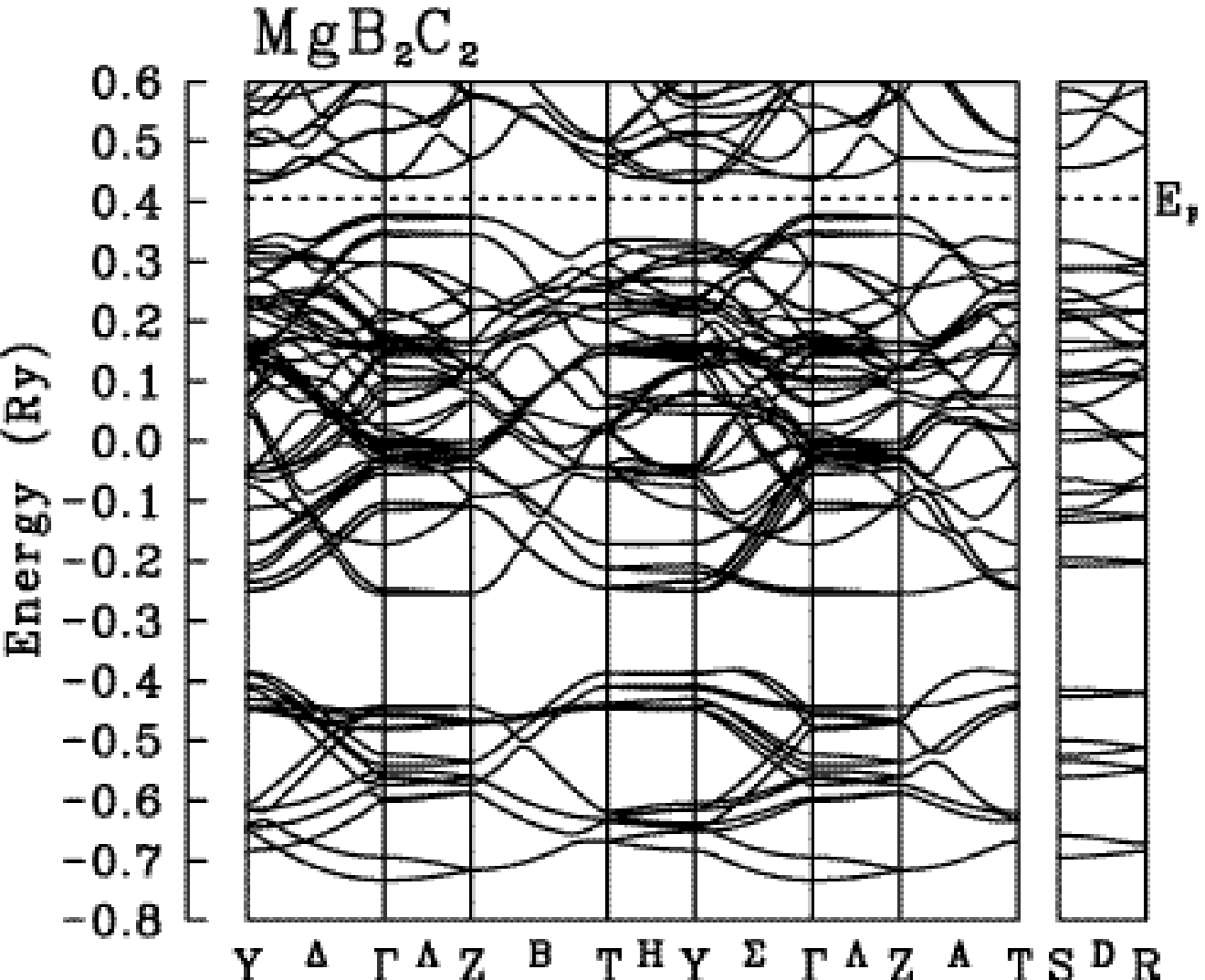}
     \caption{The calculated bandstructure for MgB$_{2}$C$_{2}$.
} 
 \label{mgb2c2band}
 \end{figure}

More interesting bandstructure is found in MgB$_2$C$_2$.
As shown in Fig.~\ref{mgb2c2band}, MgB$_2$C$_2$ becomes a semiconductor in consistent with the measured conductivity.\cite{mgb2c2cry}
The opening the gap may be due to the less anisotropic hybridyzation in the lower symmetry.
However, the dispersion of top of the valence bands are very flat along the $\Lambda$ axis, which is similar to the $\sigma$ bands in MgB$_2$.
Cylindrical Fermi surface, which is relevant in the superconductivity in MgB$_2$, can be obtained with hole doping.

\section{Lattice Constants and Structural Instability in Mg$_{1-x}$Al$_{x}$B$_2$}

With substituting Al for Mg in MgB$_2$, the superconducting transition temperature gradually decreases and structural instability appears with Al content from 10 \% to 20\%.\cite{mgalb2,china1}
After the instability, the superconducting transition temperature rapidly decreases.
Partial Al substitution for Mg means electron doping to the MgB$_2$ bandstructure, leading to vanishing $\sigma$ hole Fermi surfaces.
One can expect the structural phase instability is related to the disappearance of the $\sigma$ hole.

To reveal the structural instability, the total energy calculations with a several concentration and the lattice constant are performed.
In the calculations, the lattice constant {\bf a} is assumed to be constant, as its small Al concentration dependence ({\bf a}=3.083\AA\ in MgB$_2$ to 3.006 \AA\ in AlB$_2$) and the virtual lattice approximation is adopted, in which the atom with the atomic number of $12.x$ is used for the case of Al concentration $x$.

\begin{figure}[mgalb2]
     \centering
     \includegraphics{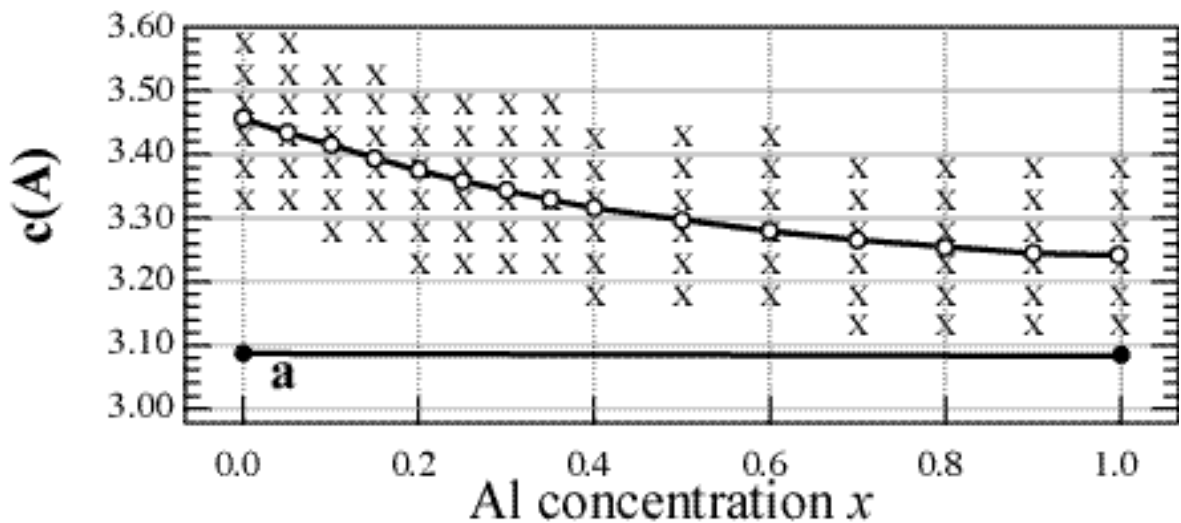}
     \caption{The optimized lattice constant {\bf c} (open curcle) vs Al-concentration $x$, keeping {\bf a} constant. The total energies are calculated at the point denoted by x.
} 
 \label{mgalb2}
\vfill
 \end{figure}

The total energies are obtained at the points in Fig.~\ref{mgalb2}, then the optimized lattice constant {\bf c} for each concentration $x$ are determined by using parabolic fitting.
The total energy as a function of {\bf c} is well reproduced as a parabolic function, so that the present calculations cannot predict the observed structural instability for any Al concentration.
There is another structural instability reported around $x=$70\%, where the superconductivity disappears.\cite{china2}
Nevertheless the optimized {\bf c} decreases monotonically with increasing $x$, as in Fig.~\ref{mgalb2}.
Concerning that Aluminum ordering is observed for $x=0.5$\cite{china2}, the virtual lattice approximation may not be applied to the system Mg$_{1-x}$Al$_{x}$B$_2$.

For AlB$_2$, with using the measured {\bf a}=3.006\AA, the optimized {\bf c} is successfully obtained as 3.245\AA\ for the experimental value 3.251\AA.
With the value {\bf a}=3.083\AA\ for MgB$_2$, which is used in Fig.~\ref{mgalb2}, {\bf c}=3.241\AA\ for AlB$_2$.
In each case, the error is less than about 0.3\% for AlB$_2$.
However, for MgB$_2$, the optimized {\bf c} is 3.457\AA\, which is 2\% smaller than the measured value 3.527\AA.
The lattice constants decrease in low temperatures, but {\bf c} is reported as 3.51504(4)\AA\ at 15K, which is only 0.3 \% smaller than in room temperature.\cite{takata}
This 2\% error should make some sense and corresponds to the jumps of two discontinuities (structural instabilities) around 20\% and 70\% Al concentration.
The larger inter-layer distance ({\bf c}) could decrease the number of the $\sigma$ holes.
Therefore there may exist some mechanism beyond the LDA that favours the small number of the $\sigma$ holes and is related to the superconductivity.
This should be solved in the future.

\section*{Acknowledgments}
The author should thank to J.~Akimitsu, K.~Ishida, Y.~Kitaoka, T.~Takahashi, H.~Kohno, K.~Miyake, M.~Imada, Y.~Kuramoto, M.~Shirai for fruitful, enlightening and stimulating discussions.
This work was partly supported by a Grant-in-Aid for Science Research from the Ministry of Education, Culture, Sports, Science and Technology, Japan.

%
%
%
%

%
%

\newpage

 \begin{table}
     \centering
     \caption{The energy and the component (\%) inside the MT spheres for the states near the Fermi energy 0.6513Ry. They consist of mainly B-$p$ component, but are extended to Mg-site then Mg-$p$ and -$d$ components appear. "--" means the component is forbidden by the symmetry.
     }
\begin{tabular}{@{\hspace{\tabcolsep}\extracolsep{\fill}}lrccccc}
\hline
\multicolumn{2}{c}{B-$p_z$($\pi$) band} &Mg-$s$&Mg-$p$&Mg-$d$&B-$s$&B-$p$\cr
\hline 
K point&0.7835Ry& -- & -- &  21& -- &  38 \cr
H point&0.5094Ry& -- &  37&   9& -- &  20 \cr
\hline
\multicolumn{2}{c}{B-$p_{xy}$($\sigma$) band} &Mg-$s$&Mg-$p$&Mg-$d$&B-$s$&B-$p$\cr
\hline
$\Gamma$ point&0.6838Ry& -- & -- &   7& -- &  55 \cr
A point&0.7070Ry& -- & -- & -- & -- &  58 \cr
\hline
     \end{tabular}
 \label{tablestates}
 \end{table}  

\newpage

 \begin{table}
     \centering
     \caption{The calculated dHvA Frequencies F in unit of Tesla and the cyclotron masses in unit of free electron mass m$_0$.
     }
\begin{tabular}{@{\hspace{\tabcolsep}\extracolsep{\fill}}lrrrrrrrr}
\hline
    &\multicolumn{2}{c}{3rd $\sigma$-band}&\multicolumn{2}{c}{4th $\sigma$-band}&\multicolumn{2}{c}{4th $\pi$-band}&\multicolumn{2}{c}{5th $\pi$-band}\\
    &F(T)& m*(m$_0$) &F(T)& m*(m$_0$)&F(T)& m*(m$_0$)&F(T)& m*(m$_0$) \cr
\hline 
H//(0001)  &1,872&0.30 &3,535 & 0.62&34,922&1.86 &31,324&1.01 \cr
           &  878&0.24 &1,889 & 0.51&      &     && \cr
H//(10\=10)&    &    &    &    & 9,486&1.00&16,449&1.00\cr
           &    &    &    &    &   524&0.29& 3,311&0.37\cr
H//(11\=20)&    &    &    &    & 4,300&0.58&2,881&0.32 \cr
           &    &    &    &    &   439&0.24&2,824&0.32 \cr
\hline
     \end{tabular}
 \label{tabledhva}
 \end{table}  

\vfill

\newpage

\bigskip
\end{document}